\documentclass[12pt,article,pra,showpacs,onecolumn]{revtex4}
\usepackage{amscd,fancyhdr}
\usepackage{amsfonts}
\usepackage[mathscr]{euscript}
\usepackage{pstricks}
\usepackage{pst-node}
\usepackage{mathrsfs}
\usepackage{float,epsfig}
\usepackage{graphicx}% Include figure files
\usepackage{dcolumn}% Align table columns on decimal point
\usepackage{bm}% bold math
\usepackage{amsmath,amssymb,amsthm}
\usepackage[colorlinks,citecolor=green,linkcolor=blue,anchorcolor=blue]{hyperref}%cite color
\usepackage[perpage]{footmisc} %footnote formate
\usepackage{indentfirst}
\makeatletter
\def\bib@device#1#2{}
\makeatother
\makeatletter
\renewcommand{\section}{\@startsection{section}{1}{0mm}
{-\baselineskip}{0.5\baselineskip}{\bf\leftline}} \makeatother
\makeatletter
\def\@hangfrom@section#1#2#3{\@hangfrom{#1#2#3}}
\makeatother
\begin{document}
\title{Hawking Radiation as tunneling and the unified first law of
thermodynamics for a class of dynamical black holes}

\author{Jiang Ke-Xia$^{a,}$\footnote{E-mail:
kexiajiang@126.com, kexiachiang@gmail.com}, KE
San-Min$^{b,}$\footnote{E-mail: ksmingre@163.com} and  PENG
Dan-Tao$^a$\footnote{E-mail: dtpeng@nwu.edu.cn}}

\affiliation{$^a$Institute of Modern Physics, Northwest University,
Xi¡¯an 710069, China}

\affiliation{$^b$College of Science, Chang'an  University, Xi'an
710064, China}

%%%%%%%%%%%%%%%%%%%%%%%%%%%%%%%%%%%%%%%%%%%%%%%%%%%%%%%%%%%%%%%%%%%%%%%%%%%%%%%%%%

%%%%%%%%%%%%%%%%%%%%%%%%%%%%%%%%%%%%%%%%%%%%%%%%%%%%%%%%%%%%%%%%%%%%%%%%%%%%%%%%%%%
\begin{abstract}
An analysis of relations between the tunneling rate and the unified
first law of thermodynamics at the trapping horizons of two kinds of
spherically symmetric dynamical black holes is investigated. The
first kind is the Vaidya-Bardeen black hole, the tunneling rate
$\Gamma \sim e^{\triangle S}$ can be obtained naturally from the
unified first law at the apparent horizon, which holds the form
$dE_{H}=TdS+WdV$. Another is the McVittie solution, the action of
the radial null geodesic of the outgoing particles does not always
has a pole at the apparent horizon, while the ingoing mode always
has one. The solution of the ingoing mode of the radiation can be
mathematically reduced to the case in the FRW universe smoothly.
However as a black hole, the physical meaning is unclear and even
puzzling.
\end{abstract}
%%%%%%%%%%%%%%%%%%%%%%%%%%%%%%%%%%%%%%%%%%%%%%%%%%%%%%%%%%%%%%%%%%%%%%%%%%%%%%%%%%
%%%%%%%%%%%%%%%%%%%%%%%%%%%%%%%%%%%%%%%%%%%%%%%%%%%%%%%%%%%%%%%%%%%%%%%%%%%%%%%%%%
%%%%%%%%%%%%%%%%%%%%%%%%%%%%%%%%%%%%%%%%%%%%%%%%%%%%%%%%%%%%%%%%%%%%%%%%%%%%%%%%%%
\maketitle
%%%%%%%%%%%%%%%%%%%%%%%%%%%%%%%%%%%%%%%%%%%%%%%%%%%%%%%%%%%%%%%%%%%%%%%%%%%%%%%%%%
\begin{flushleft}
\hspace*{1.2cm}{\bf{keyword}}: Black hole  Tunneling Apparent
horizon
\end{flushleft}
%%%%%%%%%%%%%%%%%%%%%%%%%%%%%%%%%%%%%%%%%%%%%%%%%%%%%%%%%%%%%%%%%%%%%%%%%%%%%%%%%%

\section{Introduction}\label{sec.Introduction}

To establish thermodynamics of dynamical spacetimes and to know how
it is related with gravity are important problems in General
Relativity. Understanding Hawking radiation is one of the key issues
in steps toward this aim. Since Hawking's original work
\cite{Hawking}, several derivations of Hawking radiation have been
proposed in the literatures \cite{derivations}. Recently, a
semi-classical tunneling one \cite{WK, Parikh, SSPS}, attracts many
people's attention. The main ingredient of this method is the
consideration of energy conversion in tunneling of a thin sell from
the hole. Many works \cite{Many cal.} have been investigated for
further development of this approach, and the method worked
perfectly. However, for criticism and counter criticism
see~\cite{criticism}. More recently, general analysis \cite{HZZ,
Sarkar, Pilling, ZCZ} using this method gave an interesting result:
the tunneling rate $\Gamma \sim e^{\triangle S}$ arises as a
consequence of the first law of thermodynamics for horizons holds
the form, $TdS=dE_{H}+PdV$.

However, most investigations of Hawking radiation were based on
stationary black hole spacetimes, where the globally defined surface
gravity corresponds to the Hawking temperature. Locally, it is not
clear whether there is an event horizon associated with a certain
dynamical spacetime and this causes the difficulty to discuss
Hawking radiation for a dynamical situation. Recently, Hayward
\emph{et al.} \cite{HCVNZ} proposed a locally defined Hawking
temperature for dynamical black holes where the Parikh-Wilczek
\cite{Parikh} tunneling method was used. Further, using the
tunneling method, there has been proved to be a Hawking radiation
associated with the locally defined apparent horizon of the
Friedmann-Robertson-Walker (FRW) universe \cite{Cai-Li}, where the
Hawking temperature was measured by an observer with the Kodama
vector \cite{KHMR} inside the horizon.

Even in a dynamical spacetime the tunneling method seems powerful!
This motivate us to consider: does the result obtained in~\cite{HZZ,
Sarkar, Pilling, ZCZ} is still effective in dynamical background
spacetimes? If it is true, it must cast some lights on the
thermodynamics of dynamical spacetimes. In a previous
paper~\cite{chiang}, we have successfully extended the result to the
case of the FRW universe. Based on the unified first law
$dE_{H}=TdS+WdV$ holding on the apparent horizon, we obtained the
tunneling rate naturally. The dynamical surface gravity still linked
to a Hawking temperature, which was measured by a Kodama vector.
However, for dynamical black holes, it does not seem obviously
correct. In this letter, we would like to investigate the case for
two kinds of dynamical black holes. The first is the black holes in
the Vaidya spacetime \cite{Vaidya}, which have been vastly discussed
in literatures. Another is the McVittie spacetime, which is just the
Schwarzschild black hole embedded in a dynamical background, the FRW
universe. It has often been used to study the effect of the
universe's expansion on solar system dynamics \cite{McVittie,
Krasinski}. Both of the solutions are not precisely the standard
notions of black holes, but they still have horizons which the
familiar black hole theorems seem to holds. Based on the above two
cases, Hawking radiation as tunneling from the trapping horizons
using the Hamilton-Jacobi method \cite{SSPS} were analyzed
in~\cite{CNVZZ}.

We present our analysis of the two kinds of dynamical back holes in
Sec.~\ref{sec.Vaidya} and in Sec.~\ref{sec.McVittie}, respectively.
In the last section we give out our conclusions and a brief remark.

In this Letter we take the unit convention $G = c = k =\hbar= 1$.

%%%%%%%%%%%%%%%%%%%%%%%%%%%%%%%%%%%%%%%%%%%%%%%%%%%%%%%%%%%%%%%%%%%%%%%%%%%%%%
\section{The Vaidya-Bardeen black hole}\label{sec.Vaidya}
Let us consider a spherically symmetric spacetime, which is a
typical dynamical one, the Vaidya black hole. The metric of the
4-dimensional Vaidya spacetime can be written as
\begin{equation}\label{Vaidya.metric01}
    ds^2=-e^{2 \Psi(r,v)}A(r,v)dv^2+2e^{\Phi(r,v)}dvdr+r^2d\Omega^2,
\end{equation}
where $A(r,v) = 1-2m(r,v)/r$, $r$ is the radius coordinate, $v$ is
an advanced null coordinate, and $d\Omega^2$ is the line element of
a two-dimensional unit sphere. Following~\cite{CNVZZ}, for the
special case $\Psi(r,v)=\Phi(r,v)$, we call it the Vaidya-Bardeen
metric. The metric \eqref{Vaidya.metric01} can be rewritten as
$ds^2=h_{ab}dx^adx^b+r^2d\Omega^2$, with $x^a =(v, r)$.

For dynamical black holes, we prefer to Kodama-Hayward (K-H) theory
\cite{Kodama, Hayward}, where two conserved currents can be
introduced in spherical dynamical systems. The first is the Kodama
vector $K^a=- \epsilon^{a b} \nabla_{b} r$, $\epsilon^{a b}$ denotes
the volume form. For the metric \eqref{Vaidya.metric01} we have
$K^{a}=e^{-\psi(v,r)}(\partial_{v})^{a}$, and the corresponding
conserved charge is the three dimensional volume
$V={\underset{\sigma}{\int}}K^{a}d\sigma_{a}=4 \pi r^3/3$, where
$d\sigma_{a}$ is the volume form times a future directed unit normal
vector of the space-like hypersurface $\sigma_{a}$. Another is
defined as the energy-momentum density $j^{a}=T^{a}_{b}K^{b}$ along
the Kodama vector, and the conserved charge is
$E=-{\underset{\sigma}{\int}}j^{a}d\sigma_{a}$, which is equal to
the Misner-Sharp energy. The apparent/trapping horizon\footnote{In
the generalized Vaidya spacetime, the apparent horizon is just a
kind of trapping horizon defined in K-H theory. So we don't
distinguish the two concepts in our paper. For a more detailed
discussion, we prefer the reader to~\cite{CaiCHK}.} is defined by
$h^{ab}\partial_{a}r\partial_{b}r=0$. So we have $A(r,v)=0$, which
leads to the horizon $r_{H}=r_{H}(v)=2m(v,r_{H}(v))$. According to
the definition, the Misner-Sharp energy inside the apparent horizon
$r=r_{H}$ is
\begin{equation}\label{MS.mass01}
    E_{H}=\frac{r}{2}(1-h^{ab}\partial_{a}r\partial_{b}r)|_{r=r_{H}}=\frac{r_{H}}{2}.
\end{equation}
The surface gravity associated with the Vaidya-Bardeen dynamical
horizon takes
\begin{equation}\label{surface-G.01}
    \kappa=\frac{1}{2}\nabla^{a}\nabla_{a}r|_{r=r_{H}}
    =\frac{A'(r,v)}{2}|_{r=r_{H}}=\frac{1}{2r_{H}}-\frac{m'(r_{H},v)}{r_{H}},
\end{equation}
where the prime denotes the derivative with respect to $r$.

The unified first law of thermodynamics at the apparent horizon in
theVaidya spacetime holds the form \cite{CaiCHK,Ren-Li}
\begin{equation}\label{FL.AP}
    dE_{H}=TdS+WdV,
\end{equation}
where $W=-1/2h^{ab}T_{ab}$ is the work term. For the metric
\eqref{Vaidya.metric01}, by Einstein's equations $G^{a}_{b}=8 \pi
T^{a}_{b}$, one can find $T^{v}_{v}=T^{r}_{r} =-1/(4\pi
r^2_{H})\partial m/\partial r|_{r=r_{H}}$, so we have
\begin{equation}\label{WT}
    W=-\frac{1}{2}(T^{v}_{v}+T^{r}_{r})=- T^{v}_{v}=-\frac{1}{4\pi r^2_{H}}\frac{\partial m}{\partial
r}|_{r=r_{H}}.
\end{equation}

The identity~\eqref{FL.AP} can be understood from two different
sides. In standard thermodynamics, it is a connection between two
quasi-static equilibrium states of a system, which differing
infinitesimally in the extensive variables volume, entropy, energy
by $dV$, $dS$ and $dE_H$, respectively, while having the same values
the intensive variables temperature $T$, and work density $W$. Both
of the two states are spherically symmetric solutions of Einstein
equations with the radius of horizon differing by $dr_{H}$ while
having the same source $T_{\mu \nu}$. Dynamically, it is the energy
balance under infinitesimal virtual displacements of the horizon
normal to itself. From this perspective, the identity~\eqref{FL.AP}
must be linked with conservation of energy and thus to the tunneling
process.

Corresponding the above two understanding, the whole setup can be
considered from two different sides. First, as a result of
tunneling, some matter either tunnels out or in across the horizon,
therefore the energy of the whole spacetime changes, thus the energy
attributed to the shell should be given out. Second, considering the
$s$-wave WKB approximation, the imaginary part of the action is
directly related with the Hamiltonian of tunneling particles. Thus,
the first law of thermodynamics is crucial to connect the above two
sides, energy changes of whole spacetime and the Hamiltonian of
tunneling particles.

The radial null geodesic ($ds^2=d\Omega^2=0$) near the apparent
horizon for the metric \eqref{Vaidya.metric01} is
\begin{equation}\label{null geodesic02}
    \dot{r}\equiv \frac{dr}{dv}
    =\frac{1}{2}A(r,v) e^{\psi (r,v)}\simeq e^{\psi (r_{H},v)} \kappa
    (r-r_{H}),
\end{equation}
where $\kappa$ is the surface gravity \eqref{surface-G.01}.

The imaginary part of the action for an $s$-wave outgoing positive
energy particle which crosses the horizon outwards from $r_i$ to
$r_f$ can be expressed as
\begin{align}\label{outgoing action}
    {\rm{Im}} \mathcal{S}
    &={\rm{Im}} \overset{r_{f}}{\underset{r_{i}}{\int}} dr={\rm{Im}} \overset{r_{f}}{\underset{r_{i}}{\int}}
    \overset{p_{r}}{\underset{0}{\int}}dp^{\prime}_{r}dr\nonumber\\
    &={\rm{Im}}\overset{\mathcal{H}_{f}}{\underset{\mathcal{H}_{i}}{\int}}
    \overset{r_{f}}{\underset{r_{i}}{\int}}\frac{dr}{\dot{r}}d
    \mathcal{H}=-\overset{\mathcal{H}_{f}}{\underset{\mathcal{H}_{i}}{\int}}\frac{d
    \mathcal{H}}{2T} e^{-\psi(r_{H},v)},
\end{align}
where we have used the Hamilton's equation
$\dot{r}=d\mathcal{H}/dp_{r}|_{r}$, the relation between Hawking
temperature and surface gravity, $T=\kappa/2\pi$, and a contour
integral at the pole $r=r_{H}$. Evaluating of the integral
\eqref{outgoing action}, the form of the Hamiltonian $d\mathcal{H}$
is necessary to be determined out. For this, we turn to the system,
appealing to energy conversation, to guess the form of
$d\mathcal{H}$. Since the system is explicit time dependence , the
Hamiltonian is no-longer equal to the total energy of the system.
Luckily, according to K-H theory, we still can determine out the
relation between the Hamiltonian and the total energy of the system.

The total energy of the spacetime can be expressed as
\begin{equation}\label{TE.Vaidya}
    E_{T}=\frac{r_{H}}{2}
    -{\underset{\sigma}{\int}}T^{a}_{b}K^{b}d\sigma_{a},
\end{equation}
where the first term corresponds to the energy \eqref{MS.mass01}
inside the apparent horizon, the second term corresponds to the
outside and the integration extends from the apparent horizon to
infinity. Now, we can give the energy changes between the final and
initial states of the tunneling process, which contributes to the
shell in the view of a Kodama observer. By energy conservation we
have
\begin{align}\label{energy changes01}
    d\mathcal{H}_{K}=&E^{f}_{T}(r_{H}+\delta r_{H})-E^{i}_{T}(r_{H})
    =\frac{\delta r_{H}}{2}
    -\left({{\overset{\infty}{\underset{r_{H}+\delta r_{H}}{\int}}}
    -{\overset{\infty}{\underset{r_{H}}{\int}}}}
    \right) T^{v}_{v} K^{v} d \sigma_{v}\nonumber\\
    =&dE_{H}+T^{v}_{v}d V=dE_{H}-W d V,
\end{align}
where $W$ is the work term \eqref{WT}.

Since the energy difference $d\mathcal{H}_{K}$ is measured by a
Kodama observer inside the apparent horizon, in our case near the
apparent horizon, we have
\begin{equation}\label{our observer}
    d\mathcal{H}=\frac{d\mathcal{H}_{K}}{e^{-\psi(v,r)}|_{r=r_{H}}}
    =e^{\psi(v,r_{H})}d\mathcal{H}_{K}.
\end{equation}
Substituting \eqref{energy changes01} \eqref{our observer}  into
\eqref{outgoing action}, one can obtain
\begin{equation}\label{action relation}
    {\rm{Im}} \mathcal{S}=-\overset{\mathcal{H}_{f}}{\underset{\mathcal{H}_{i}}{\int}}\frac{d
    \mathcal{H}}{2T}
    =-\int\frac{dE_{h}-W d V}{2T}.
\end{equation}
Using the first law of thermodynamics \eqref{FL.AP} on the apparent
horizon, from \eqref{action relation} we finally have
\begin{equation}\label{finally equation}
    {\rm{Im}} \mathcal{S}=-\int\frac{dS}{2}.
\end{equation}
Now, one can immediately have the semi classical tunneling rate from
the Vaidya-Bardeen black hole, $\Gamma \sim e^{-2 \texttt{Im}
\mathcal{S}}=e^{\int^{S_{f}}_{S_{i}}dS }=e^{+\triangle S}$, with
$\triangle S= S_{f}-S_{i}$. This is the well-known result obtained
in~\cite{Parikh} for a general, stationary, asymptotically flat,
spherically symmetric background. And as a consequence of the first
law of thermodynamics, this result appeared in the discussion of a
static, spherically symmetric spacetime in~\cite{Sarkar, Pilling}.
Here, we have recovered it in a background of dynamical spacetime,
the Vaidya-Bardeen black hole.

%%%%%%%%%%%%%%%%%%%%%%%%%%%%%%%%%%%%%%%%%%%%%%%%%%%%%%%%%%%%%%%%%%%%%%%%%%%%%
\section{The McVittie solution} \label{sec.McVittie}
In this section, we will analyze another dynamical black holes, the
McVittie solution. As we will see below, preferring to the K-H
theory, the radiation of this kind of dynamical black hole is
puzzling. In 4-dimensional spacetime the metric of the McVittie
solution is given by \cite{C.J.Gao}
\begin{equation}\label{McVittie.metric}
    ds^2=-A(r,t)dt^2+B(r,t)(dr^2+r^2d\Omega^2),
\end{equation}
where
\begin{align}\label{metric.component}
    A(r,t)=\left [ \frac{1-(\frac{m}{a(t)r})}{1+( \frac{m}{a(t)r})}\right ]^2,
    B(r,t)=a^2(t)\left [ 1-\left ( \frac{m}{a(t) r}\right )\right ]^2.
\end{align}
When $m=0$, the metric \eqref{McVittie.metric} reduces to a flat FRW
solution with the scale factor $a(t)$; while when $a(t)= 1$ it
becomes a Schwarzschild metric with mass $m$. Takeing the so-called
Nolan gauge \cite{Nolan}, the metric \eqref{McVittie.metric} can be
expressed as
\begin{equation}\label{Nolan.metric}
    ds^2=-(A_{s}-H^{2}(t)
    r^2)dt^{2}+A^{-1}_{s}dr^{2}-2A^{-1/2}_{s}H(t)r dr dt+r^2
    d\Omega^{2},
\end{equation}
where $r\in (2m,\infty)$, $A_{s}\equiv 1-2m/r$, and $H(t)=\dot{a}/a$
is the Hubble parameter.

The energy momentum tensor of the homogeneous perfect fluid takes
\begin{equation}\label{EM.McVitti}
    T^{\mu}_{\nu}={\rm diag}(\rho, -p, -p, -p ),
\end{equation}
where $\rho = \rho(t)$ is the energy density, and $p = p(t)$ is the
pressure of the perfect fluid. The Einstein-Friedmann equations read
\begin{equation}\label{EF.eq}
    3H^2=8\pi \rho, \quad 2A^{-1/2}_{s}\dot{H}+3H^2=-8\pi p.
\end{equation}
The metric \eqref{Nolan.metric} can be rewritten as
$ds^2=h_{ab}dx^{a}dx^{b}+r^2d\Omega^{2}$, with $x^a=(t,r)$. Here the
McVittie black hole is really a fake dynamical one since we let the
mass $m=\rm{const}$. However, it is still has dynamical horizons.

The apparent/trapping\footnote{Here we still not distinguish the two
concepts, since in our calculation, using the definition of the
apparent horizon, which is equal to the trapping horizon exactly
from the K-H theory in~\cite{CNVZZ} } horizon can be given by
$h^{ab}\nabla_{a}r \nabla_{b}r =0$, and using the metric
\eqref{Nolan.metric} we have the relation
\begin{equation}\label{relation}
A_{s}|_{r_{H}}\equiv 1-\frac{2m}{r_{H}}=H^{2}(t)r^{2}_{H},
\end{equation}
which determines the radius of the apparent horizon
$r_{H}=r_{H}(t)$. The Misner-Sharp mass inside the horizon takes
\begin{equation}\label{MS.mass02}
    E_{H}=\frac{r}{2}(1-h^{ab}\partial_{a}r\partial_{b}r)|_{r=r_{H}}
    =\frac{r_{H}}{2}=m+\frac{1}{2}H^{2}(t)r^{3}_{H}.
\end{equation}
For the metric \eqref{Nolan.metric}, the surface gravity is
$\kappa=\frac{1}{2}\nabla^{a}\nabla_{a}r|_{r=r_{H}}=m/r^2_{H}-H^2
r^2_{H} -\dot{H}/(2H)$. Using \eqref{relation} it can be expressed
as
\begin{equation}\label{surface-G.02}
    \kappa=\frac{3m}{r^2_{H}}-\frac{1}{r_{H}}\left ( 1-\frac{3H^2r^2_{H}-1}{4
    H^2r^2_{H}}\dot{r}_{H}\right ).
\end{equation}
One can easily see when $a(t)=\rm const$, it is just the surface
gravity of the Schwarzschild black hole at the even horizon, which
takes $\kappa=1/(2r_{H})$ with $r_{H}=2m$. While when $m=0$, it
reduce to the surface gravity of a flat FRW universe at the apparent
horizon $r_{H}=1/H(t)$, which is $\kappa= -\left ( 1-\dot{r}_{H}/(2
H r_{H})\right )/r_{H}$ \cite{Cai03}.

Since we prefer to the K-H theory, we assume that the unified first
law of thermodynamics at the apparent/trapping horizon of the
McVittie black hole still takes~\cite{CNVZZ}
\begin{equation}\label{FL.AP02}
    dE_{H}=TdS+WdV,
\end{equation}
with the work term $W=(\rho-p)/2$, and $S=\pi r^2_{H}$. We interpret
$S$ as the entropy inside the apparent/trapping horizon. A more
detailed discussion is indeed need, however, it's beyond our present
paper.

The radial null geodesic for the metric \eqref{Nolan.metric} is
\begin{equation}\label{Nolan.null.geodesic}
    \dot{r} \equiv \frac{dr}{dt}
    =H r \sqrt{A_{s}}\pm A_{s} ,
\end{equation}
where $+/-$ corresponding the outgoing/ingoing positive energy
particles respectively. However, from \eqref{Nolan.null.geodesic}
one can see that for the outgoing mode, the action at the apparent
horizon $r=r_{H}$, does not always has a pole, unless the expansion
rate is slowly enough ($H=\dot{a}/a \simeq 0$). In that case it
reduces to a Schwarzschild one, which the Hawking radiation has been
vastly discussed in literatures. While the action of the ingoing
mode always has a pole at the horizon. So, from the view of
tunneling, the McVittie solution is more of a FRW universe than a
Schwarzschild black hole.

In the following, we would like to investigate the tunneling process
of the ingoing particles. Using \eqref{relation} and
\eqref{surface-G.02}, near the horizon $r_{H}$ , the equation
\eqref{Nolan.null.geodesic} can be rewritten as
\begin{align}\label{Nolan.null.geodesic02}
    \dot{r}&\simeq-\left ( \frac{3m}{r^2_{H}}-\frac{1}{r_{H}}\right
    )(r-r_{H})\nonumber\\
    &= -\left [1+\frac{3H^2 r^2_{H}-1}{4H^2r^3_{H}}
    \left ( \frac{3m}{r^2_{H}}-\frac{1}{r_{H}}\right
    )^{-1}\dot{r}_{H}\right ]^{-1}\kappa (r-r_{H}).
\end{align}
Investigating an ingoing mode, we hope the McVittie spacetime can be
smoothly reduced to the FRW universe. Assuming that the mass $m$ is
small enough, so we have $\kappa \simeq -\left ( 1-\dot{r}_{H}/(2 H
r_{H})\right )/r_{H} < 0$ \cite{Cai03}.

Similar with the procedure in the Sec. \ref{sec.Vaidya}, the
imaginary part of the action for an $s$-wave ingoing positive energy
particle which crosses the horizon inwards from $r_i$ to $r_f$ takes
\begin{align}\label{ingoing action}
    {\rm{Im}}\mathcal{ S}
    =\overset{\mathcal{H}_{f}}{\underset{\mathcal{H}_{i}}{\int}}\frac{d
    \mathcal{H}}{2T} \left [1+\frac{3H^2
    r^2_{H}-1}{4H^2r^3_{H}}\left ( \frac{3m}{r^2_{H}}-\frac{1}{r_{H}}\right
    )^{-1}\dot{r}_{H}\right ],
\end{align}
here, the relation between Hawking temperature and surface gravity
is $T=|\kappa|/2\pi$.

Corresponding the metric \eqref{Nolan.metric}, the Kodama vector is
$K^{a}=(\partial_{t})^{a}$. Interestingly, the Kodama vector is
accidentally equal to the timelike Killing vector in stationary
black hole systems. So, from this point of view, the radiation is
more of a stationary Schwarzschild black hole than the dynamical FRW
universe. As we have analyzed in Sec. \ref{sec.Vaidya}, in this
situation, the total energy of spacetime still takes $E_{T}=r_{H}/2
-{\underset{\sigma}{\int}}T^{a}_{b}K^{b}d\sigma_{a}$. The energy
contributing to the shell in the view of a Kodama observer is
\begin{align}\label{energy changes}
    d\mathcal{H}_{K}
    =&E^{f}_{T}(r_{H}+\delta r_{H})-E^{i}_{T}(r_{H})
    =\frac{\delta r_{H}}{2}
    -\left({{\overset{\infty}{\underset{r_{H}+\delta r_{H}}{\int}}}
    -{\overset{\infty}{\underset{r_{H}}{\int}}}}
    \right) T^{v}_{v} K^{v} d \sigma_{v}\nonumber\\
    =&dE_{H}-\rho dV.
\end{align}
Since the Kodama vector is equal to the timelike Killing vector, we
have $d\mathcal{H}=d\mathcal{H}_{K}$. From \eqref{ingoing action}
and \eqref{energy changes} one obtains
\begin{equation}\label{ingoing.action02}
    {\rm{Im}} \mathcal{S}
    ={\int}\frac{dE_{H}-\rho dV}{2T} \left [1+\frac{3H^2
    r^2_{H}-1}{4H^2r^3_{H}}\left ( \frac{3m}{r^2_{H}}-\frac{1}{r_{H}}\right
    )^{-1}\dot{r}_{H}\right ].
\end{equation}
The integrated term of the above equation can be further simplified.
Using \eqref{EF.eq} and \eqref{MS.mass02}, we get
\begin{align}\label{energy.relation}
    dE_{H}&=H\dot{H}r^3_{H}dt+\frac{3}{2}H^2r^2_{H}dr_{H}
    =\frac{4\pi}{3}r^3_{H}\dot{\rho}dt+\frac{3H^2}{8\pi}dV\nonumber\\
    &=Vd\rho+\rho
    dV=d(\rho V).
\end{align}
Combining \eqref{EF.eq}, \eqref{relation}, and
\eqref{energy.relation}, we have
\begin{align}\label{reduce}
    &(dE_{H}-\rho dV)\frac{3H^2
    r^2_{H}-1}{4H^2r^3_{H}}\left ( \frac{3m}{r^2_{H}}-\frac{1}{r_{H}}\right)^{-1}\dot{r}_{H}\nonumber\\
    &=Vd\rho\frac{1}{2H^2r^2_{H}}\dot{r}_{H}=-\frac{4}{3}\pi \dot{\rho}\frac{r_{H}}{2H^2}dr_{H}=-\frac{\dot{H}r_{H}}{2H}dr_{H}\nonumber\\
    &=2\pi (\rho+p)
    r^2_{H}dr_{H}=\frac{1}{2}(\rho+p)dV.
\end{align}
Substituting \eqref{reduce} into \eqref{ingoing.action02}, one have
\begin{equation}\label{ingoing.action03}
    {\rm{Im}} \mathcal{S}
    ={\int}\frac{dE_{H}-W dV}{2T}
\end{equation}
where $W=(\rho-p)/2$ is the work term. Using the first law of
thermodynamics \eqref{FL.AP02} on the horizon $r_{H}$, the semi
classical tunneling rate takes $ \Gamma \sim e^{-2
\texttt{Im}\mathcal{ S}}=e^{-\int^{S_{f}}_{S_{i}}dS }=e^{-\triangle
S}$ with $\triangle S= S_{f}-S_{i}$. Indeed, the result can be
mathematically reduced to the case of the FRW universe \cite{chiang}
smoothly. However, as a black hole, the physical meaning of the
ingoing mode radiation of the McVittie spacetime is puzzling.
%%%%%%%%%%%%%%%%%%%%%%%%%%%%%%%%%%%%%%%%%%%%%%%%%%%%%%%%%%%%%%%%%%%%%%%%%%%%%
\section{Conclusion and Remarks}\label{sec.Conclusion}
In this Letter preferring to the K-H theory, we have extended the
work \cite{HZZ, Sarkar, Pilling} to investigate two kinds of
dynamical black holes, the Vaidya-Bardeen black hole and the
McVittie black hole. In the Vaidya-Bardeen spacetime, the tunneling
rate $\Gamma \sim e^{\triangle S}$ can be obtained naturally from
the unified first law at the apparent horizon, which holds the form
$dE_{H}=TdS+WdV $. In the McVittie case, we find the action of the
radial null geodesic of the outgoing particles does not always has a
pole at the apparent horizon, while the ingoing mode always has one.
From the view of tunneling, the McVittie black hole is more of the
FRW universe than a Schwarzschild black hole. Assuming the mass $m$
is small enough, the McVittie solution can be reduced to the FRW
universe smoothly. The tunneling rate of ingoing particles still can
be obtained from the unified first law holds on the apparent
horizon, where the procedure mathematically resembles with the case
in the FRW universe \cite{chiang}. However, as a black hole, the
physical meaning of this kind radiation is unclear and even
puzzling. In this sense, the McVittie spacetime may also not
actually be viewed as a dynamical black hole, despite its
resemblance.

\section*{Acknowledgments} The author(K.-X. Jiang) thank Dr. Cheng-Yi Sun for
kind help and useful discussion. This work is fund by National
Natural Science Foundation of China (Grant No. 10875060), and the
Natural Science Foundation of Shaanxi Education Bureau of China
(Grant No. 07JK394).

%% The Appendices part is started with the command \appendix;
%% appendix sections are then done as normal sections
%% \appendix

%% \section{}
%% \label{}

\end{document}